\documentclass[prd,twocolumn]{revtex4}
\usepackage{xcolor}
\usepackage[utf8]{inputenc}
\usepackage{graphicx}
\usepackage{mhchem}
\usepackage{braket}
\usepackage{float}
\usepackage{amssymb}
\usepackage{amsmath}
\usepackage[titletoc]{appendix}
\usepackage{siunitx}
\usepackage{hyperref}
\hypersetup{colorlinks=true, linkcolor=magenta, citecolor=magenta, urlcolor=magenta}
\usepackage{ulem}

\newcommand\blue[1]{{\color{blue}#1}}

\begin{document}
\title{\blue{Evidence of controlling vortex matter via a superconducting Nanobridge}}
\author{C. A. Aguirre$^{1}$\href{https://orcid.org/0000-0001-8064-6351}{\includegraphics[scale=0.05]{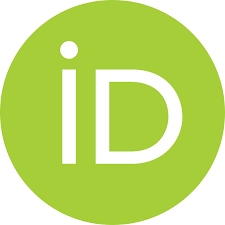}}, J. Faúndez$^2$\href{https://orcid.org/0000-0002-6909-0417}{\includegraphics[scale=0.05]{orcid.png}}
P. Díaz$^{3}$\href{https://orcid.org/0000-0002-8624-1707}{\includegraphics[scale=0.05]{orcid.png}}, D. Laroze$^{4}$\href{https://orcid.org/0000-0002-6487-8096}{\includegraphics[scale=0.05]{orcid.png}},  A. S. Mosquera Polo$^{5}$\href{https://orcid.org/0000-0002-4270-6883}
{\includegraphics[scale=0.05]{orcid.png}}, N. C. Costa$^{2}$\href{https://orcid.org/0000-0003-4285-4672}
{\includegraphics[scale=0.05]{orcid.png}}, J. Barba-Ortega$^{6,7}$\href{https://orcid.org/0000-0003-3415-1811}{\includegraphics[scale=0.05]{orcid.png}}}
\affiliation{$^1$ Departamento de F\'isica, Universidade Federal de Mato-Grosso, Cuiab\'a, Brasil.}
\affiliation{$^2$ Instituto de F\'sica, Universidade Federal do Rio de Janeiro Cx.P. 68.528, 21941-972 Rio de Janeiro RJ, Brazil}
\affiliation{$^3$ Departamento de Ciencias F\'isicas, Universidad de La Frontera, Temuco, Casilla 54-D, Chile.}
\affiliation{$^4$ Instituto de Alta Investigación, Universidad de Tarapac\'a, Casilla 7D, Arica, Chile.}
\affiliation{$^5$ Facultad de Ingenieria, Universidad del Magdalena, Santa Marta, Colombia.}
\affiliation{$^6$ Departamento de F\'isica, Universidad Nacional de Colombia, Bogot\'a, Colombia.}
\affiliation{$^7$ Foundation of Researchers in Science and Technology of Materials, Bucaramanga, Colombia.}
\date{\today}
\begin{abstract} 
We theoretically investigate the magnetic response on a three-dimensional superconducting nanobridge system, which is compound of two parallel parallelepiped (samples) connected through a nanobridge of size $\mathbf{L}$ and thickness $\mathbf{x}$,  which mediates interactions between them. This study is conducted in the presence of a magnetic field $\mathbf{H}$ and the transport of a direct current $\mathbf{J}$. We use the well-know time dependent Ginzburg-Landau theory ($\mathbf{TDGL}$) for analyzed the possible effects on the density Gibbs free energy $\mathbf{F}$, magnetization $\mathbf{M}$, and superconducting electronic Cooper pair density $|\psi|^{2}$. We are interested in studying two cases: varying the \(\mathbf{L}\) and \(\mathbf{x}\) of the nanobridge in the absence of induced \(\mathbf{J}\), and including the induction of external \(\mathbf{J}\) for fixed \(\mathbf{L}\) and \(\mathbf{x}\). We find that \(\mathbf{L}\) and \(\mathbf{x}\) play an essential role in stabilizing (controlling) vortex states in the nanobridge, and the presence of induced \(\mathbf{J}\) ($\mathbf{J}>0$ and $\mathbf{J}<0$), with a fixed \(\mathbf{L}\) and \(\mathbf{x}\),  causes the movement of vortex states in the nanobridge just when $\mathbf{J}$ is induced at both faces of superconducting nanobridge system.\\\\

\end{abstract}
\maketitle
\section{Introduction}\label{sec1}
In recent years, there has been renewed interest in studying both conventional and unconventional superconductors, driven by the quest to achieve superconductivity at ambient temperature and pressure, often considered the "holy grail" of condensed matter physics \cite{In1,In2,In3}. Over the past decades, various theories have been developed to describe different aspects of superconductivity. These theories have expanded to address multi-band systems, non-monotonic interactions (such as short-range repulsion and long-range attraction), and the coexistence of diverse states of matter, including ferromagnetism and diamagnetism \cite{In3}. Furthermore, numerous new states have been discovered and measured in laboratories worldwide in recent years \cite{In4,In5,In6}. A significant focus in the current development and study of superconducting phenomena is on mesoscopic or nanoscale systems, which can exhibit various geometries and confinements \cite{In4,In6}. In materials physics and quantum engineering, superconducting nanostructures have become crucial due to their unique properties and potential applications in diverse fields, such as medicine, biology, and as a foundation for future technological advances \cite{In7,In8,In9,In92,In10}.\\\\
One of the possibilities in these nanosystems is the use of a superconducting nanobridge, a small structure that connects two superconducting sections or samples. When cooled below a critical temperature $T_{c}$, this nanobridge allows electrical current to flow without resistance. Additionally, it may enable the flow of super-electrons and potentially vortex states between the main sections. This phenomenon, of quantum confinement, not only challenges our conventional understanding of electrical conduction and the possibility of vortex dynamics, but also opens the door to technological innovations in areas such as quantum computing, superconducting quantum interference devices (SQUID) and ultrasensitive sensors, for the measurement of weak magnetic fields \cite{In92,In10}. These tiny bridge possibilities not only maintain the zero-resistance property, essential property of superconductivity states, but can also manifest the Josephson effect, where Cooper pairs tunnel through the bridge, allowing voltage-free currents and the real possibility of fluxoid motion. Furthermore, the quantum coherence maintained throughout superconducting nanobridges is essential for the use in emerging quantum technologies, as the basis of \textit{qubits}, the building blocks of quantum computers, taking advantage of their superposition and quantum entanglement capabilities, in conjunction with the manipulation (Controlling) \cite{In11,In13,In14}. In addition, it is with this that the control of vortex states in mesosystems opens a door to superconducting electronic applications of different types, which are important for the evolution of superconducting technology and developments of new mesoprocessors, with greater capacity, efficiency, and computing speed \cite{In18,In19}.\\\\ 
In this article, we propose a study of a superconducting nanobridge, which couples two superconducting samples, using the well-established time-dependent Ginzburg-Landau ($\mathbf{TDGL}$) theory. We consider a nanobridge with  length $\mathbf{x}$ and the thickness $\mathbf{L}$.
Subsequently, modifications are introduced on the samples by applying an external magnetic field, which generates a current described by the Ampere-Maxwell equation on one of the lateral faces of the superconducting nanobridge system (see Fig. \ref{Layout}(b)). This setup enables the analysis of the interaction between the generated vortex states and the Lorentz force across the nanobridge. In this study, positive and negative currents are included as a control mechanism for the entry of fluxoids. Finally, currents are applied to both lateral faces of the samples to investigate  the generation of vortex states, as well as the potential interaction or control between the superconducting samples.\\\\
This work is organized as follows. In Section \ref{sec2}, we detail the theoretical framework used to study the superconducting electronic parameters, including magnetization, Gibbs free  energy, and vortex states. Section \ref{sec3}, (\ref{sec3a} and \ref{sec3b}) provide a discussion of the results obtained from numerical simulations of the non-linear partial differential Ginzburg-Landau equations. In Section \ref{sec4}, we summarize the main conclusions of our study.
\section{Model and methodology}\label{sec2}
\begin{figure}
\centering   
\includegraphics[scale=0.65]{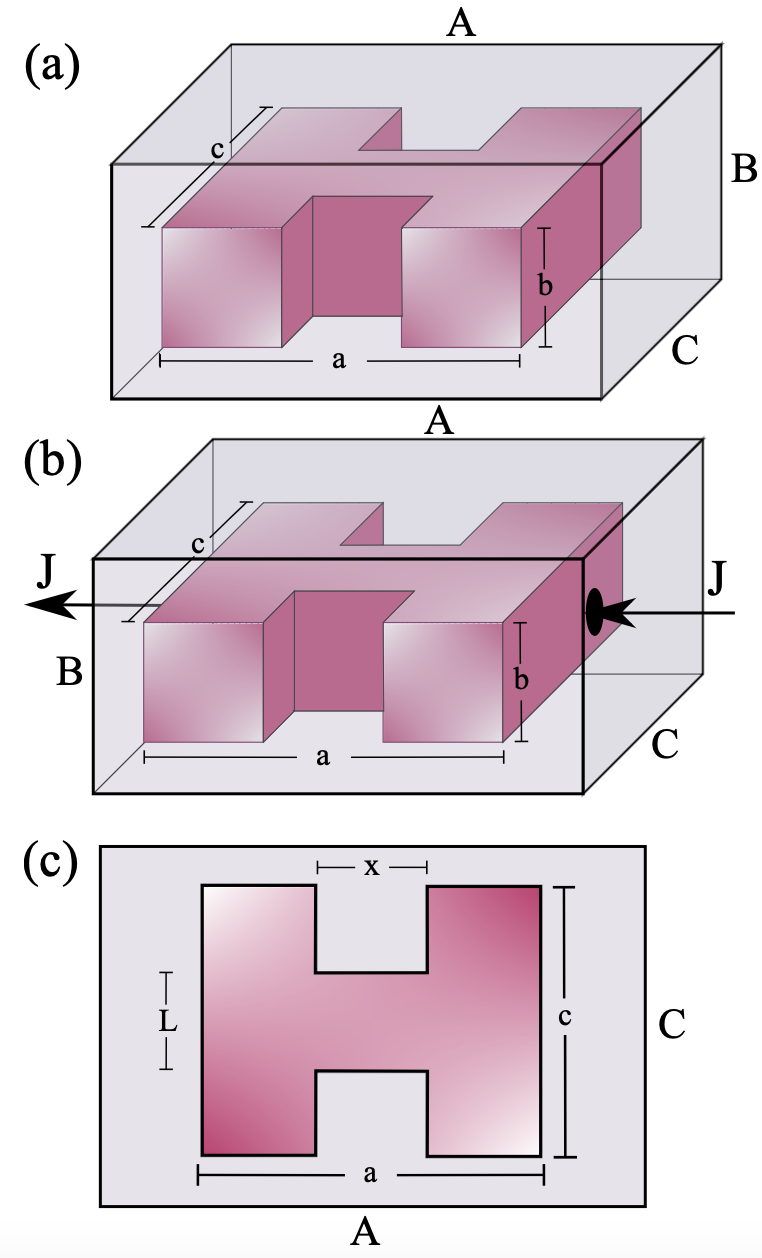}
\caption{Sketch of the superconducting nanobridge system. The dimensions of the external numerical mesh are ${\bf A} \times {\bf B} \times {\bf C}$ and the dimensions of the real sample are ${\bf a} \times {\bf b} \times {\bf c}$, where ${\bf A} = {\bf B} = 20\xi$, ${\bf C} = 10\xi$, ${\bf a} = {\bf b} = 10\xi$, and ${\bf c} = 5\xi$. (a) The real three-dimensional superconducting nanobridge system studied. (b) The superconducting nanobridge system with the inclusion of external currents ${\bf J}$. (c) The projection of the superconducting sample with the definitions of ${\bf x}$ (the distance between the superconducting side samples) and ${\bf L}$.}
\label{Layout}
\end{figure}

We studied the vortex state and the magnetic response in a superconducting three-dimensional nanobridge in an external magnetic field, ${\bf H}$, applied in the $z$-direction. The geometry of the problem that we investigate is illustrated in Fig. \ref{Layout}(a-c). The domain $\Omega_{sc}$ is filled by the mesoscopic superconducting nanobridge of high $C$ and base of size $\mathbf{A} \times \mathbf{B}$. The interface between this region and the vacuum is denoted by $\partial\Omega_{sc}$. Due to demagnetization effects, we consider a larger domain $\Omega$ of dimensions $\mathbf{A}\times \mathbf{B} \times \mathbf{C}$, such that $\Omega_{sc}\subset\Omega$. The superconducting interface is indicated by $\partial\Omega$. The domain $\Omega$ is taken sufficiently large such that the local magnetic field equals the applied field ${\bf H}$ at the surface $\partial\Omega$ (see reference \cite{In15,In16,In17} for more details). 

The general form of {\bf TDGL} equations in dimensionless units are given by \cite{In18,In19,In20,In21}:\\
\begin{eqnarray}
\frac{\partial \psi}{\partial t}  &=&   -(-i\mbox{\boldmath $\nabla$}-{\bf A})^{2}\psi+\psi(1-|\psi|^2)\;,\;\;\;{\rm in}\;\Omega_{sc}\;,  \label{EQ4} \\ \nonumber\\ 
\frac{\partial{\bf A}}{\partial t} &=& 
\left \{ 
\begin{array}{ll}
{\bf J}_s-\kappa^2\mbox{\boldmath $\nabla$}\times\mbox{\boldmath $\nabla$}\times
{\bf A}\;,\;\;\;{\rm in}\;\Omega_{sc}\;, \\
-\kappa^2\mbox{\boldmath $\nabla$}\times\mbox{\boldmath $\nabla$}\times
{\bf A}\;,\;\;\;{\rm in}\;\Omega\backslash\Omega_{sc}\;,
\end{array}
\right .\;, \label{EQ5}
\end{eqnarray}
where:
\begin{equation}
{\bf J}_s = {\rm Re}\left [
\bar{\psi}{(-i\mbox{\boldmath $\nabla$}-{\bf A})}
\psi\right]
\label{EQ6}
\end{equation}
${\bf J}_s$ correspond to the superconducting current density. In Eqs.~(\ref{EQ4})-(\ref{EQ6}) dimensionless units were introduced as follows: the order parameter $\psi$ is in units of $\psi_{\infty}=\sqrt{-\alpha/\beta}$, where $\alpha$ and $\beta$ two phenomenological constants that depends on the material; Gibbs free energy $F$ in units of $F_0= H^2_{c2}/4\pi\xi$, lengths are in units of the coherence length $\xi$; time is in units of Ginzburg-Landau characteristic time $t_{GL}=\pi\hbar/8K_BT_c$; fields are in units of $H_{c2}$, where $H_{c2}$ is the bulk second critical field; the potential vector ${\bf A}$ is in units of $\xi H_{c2}$; $\kappa=\lambda/\xi$ is the Ginzburg-Landau parameter ($\mathbf{GLP}$).\\\\
For the numerical simulations, we employ the $\mathbf{U}$-$\psi$ method (link variables method), which has been extensively studied with various modifications both with and without currents, across different configurations and dimensionalities \cite{Rev1,Rev2,Rev3,Rev4}. These studies include interactions among multiple bands, enabling the description of unconventional vortex states in superconducting samples \cite{Rev5,Rev6,Rev7} with the well know time-converge rule (Courant-Friedrich-Levy rule) \cite{Rev3,Rev4,Rev5}:\\
\begin{eqnarray}
\Delta t \leq min \bigg\{\frac{a^2}{4}, \frac{\beta a^2}{4\kappa^2}\bigg\}; \qquad a^2 = \frac{2}{\frac{1}{\delta x^2} + \frac{1}{\delta y^2}+\frac{1}{\delta z^2}},
\end{eqnarray}\\
where the time is in units of the Ginzburg-Landau characteristic time $t_{GL} = \pi \hbar / 8k_B T_{c1}$, $H_{c2} \xi_{10}$, $\delta x = \delta y = \delta z = 0.1$ and $\mathbf{GLP}$    $\kappa=1.0$. In addition, to include the induced currents, we modify the magnetic boundary conditions to generate induced current densities just at the boundaries. On the other hand, for the tolerance on the convergence rule, for the order parameter $\psi$, we use $\epsilon=1.0^{-9}$, guaranteeing slight differences between the values obtained in each convergence iteration loop, and due to the scheme used (Crank-Nicholson), errors are obtained of the order  $O(\delta x)^{2}$ for the space and time. Finally, for general boundary conditions, we use the Newman's boundary condition ${\bf n} \cdot (i\mbox{\boldmath $\nabla$} + {\bf A}) \psi = 0$, with ${\bf n}$ being a surface normal vector \cite{Demler2018,Kalashnikov2023,Metlitski2017,Singh2021,Zolotov2014,Levchenko2020}.

We begin our investigation by considering the general form of the $\mathbf{TDGL}$ equations as applied to the superconducting nanobridge system. Our focus is on two primary cases: (i) the absence or (ii) the presence of induced currents. For the latter, two sub-cases are considered, one with a single lateral current, and another with the inclusion on both sides of the superconducting nanobridge system.\newpage
\section{Numerical Results}\label{sec3}

\subsection{Superconducting nanobridge system with absence of induced external currents}\label{sec3a}
\begin{figure}
    \centering
    \includegraphics[scale=0.7]{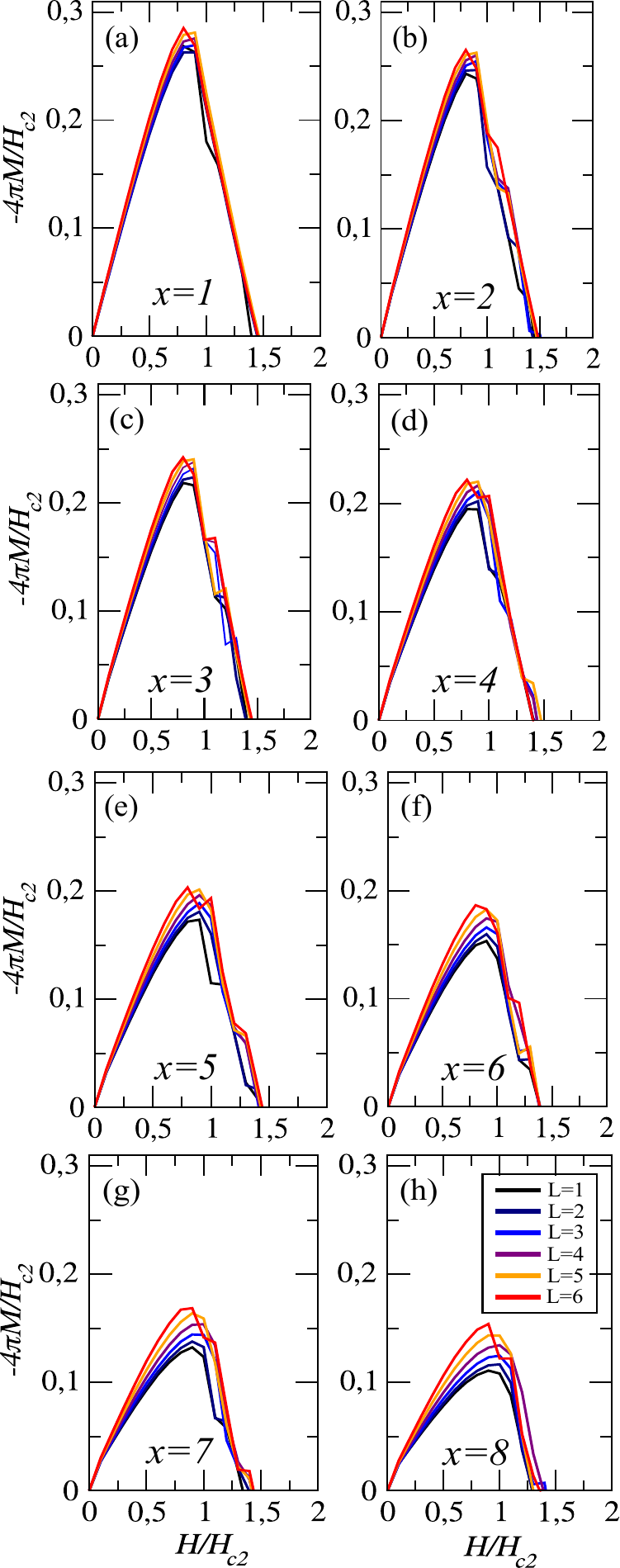}
    \caption{(a)-(h) Magnetization, $-4\pi\mathbf{M}/\mathbf{H}_{c2}$, as a function of the external magnetic field, $\mathbf{H}/\mathbf{H}_{c2}$, for  $1\leq \mathbf{L} \leq 6$ ($\Delta \mathbf{L}=1$), and $1 \leq \mathbf{x} \leq 8$ ($\Delta \mathbf{x}=1$).}
    \label{Magnetizacion1}
\end{figure}

In the absence of external current, \textbf{J}, we examine the behavior of $-4\pi\mathbf{M}/\mathbf{H}_{c2}$,  $\mathbf{F}/\mathbf{F}_{0}$, and  $|\psi|^2$ for the nanobridge superconducting system depicted in Fig. \ref{Layout}(a). Figs. \ref{Magnetizacion1}(a)-(h) show  $-4\pi\mathbf{M}/\mathbf{H}_{c2}$ as a function of $\mathbf{H/H_{c2}}$ for the nanobridge superconducting  with varying separations between the lateral faces sections ($\mathbf{x}$= bridge length), $1\leq \mathbf{x}\leq8$, while keeping  a fixed $\mathbf{L}$.
In the regime of $\mathbf{H}/\mathbf{H_{c2} } \lesssim 1.0$, the  $-4\pi \mathbf{M}/\mathbf{H}_{c2}$ increases linearly until it reaches a peak. This behavior indicates that the nanobridge superconducting system remains in the Meissner-Ochsenfeld state, where $\mathbf{H}/\mathbf{H}_{c2}$ is effectively shielded or expelled. As the bridge thickness $\mathbf{L}$ increases, and for all values of $\mathbf{x}$, we observe minor variations in the maximum magnetization, suggesting that $\mathbf{L}$ plays a significant role in stabilizing of $-4\pi \mathbf{M}/\mathbf{H}_{c2}$. For $\mathbf{H}/\mathbf{H}_{c2} \gtrsim 1.0$, $-4\pi \mathbf{M}/\mathbf{H}_{c2}$ decreases linearly for all values of $\mathbf{x}$ and $\mathbf{L}$ until it reaches a value of $\mathbf{H}/\mathbf{H}_{c2} \sim 1.5$. At this point, there is an absence of magnetization in the system, indicating that the nanobridge superconducting system is in the normal state. Furthermore, as $\mathbf{L}$ increases, the $-4\pi \mathbf{M}/\mathbf{H}_{c2}$ maximum shows a strong dependence on $\mathbf{x}$, with a maximum of $-4\pi \mathbf{M}/\mathbf{H}_{c2} \sim 0.29$ at $\textbf{x} \sim 1$ and decreasing to $-4\pi \mathbf{M}/\mathbf{H}_{c2} \sim 0.15$ at $\mathbf{x} = 6$.\\\\\  

\begin{figure}
    \centering
    \includegraphics[scale=0.56]{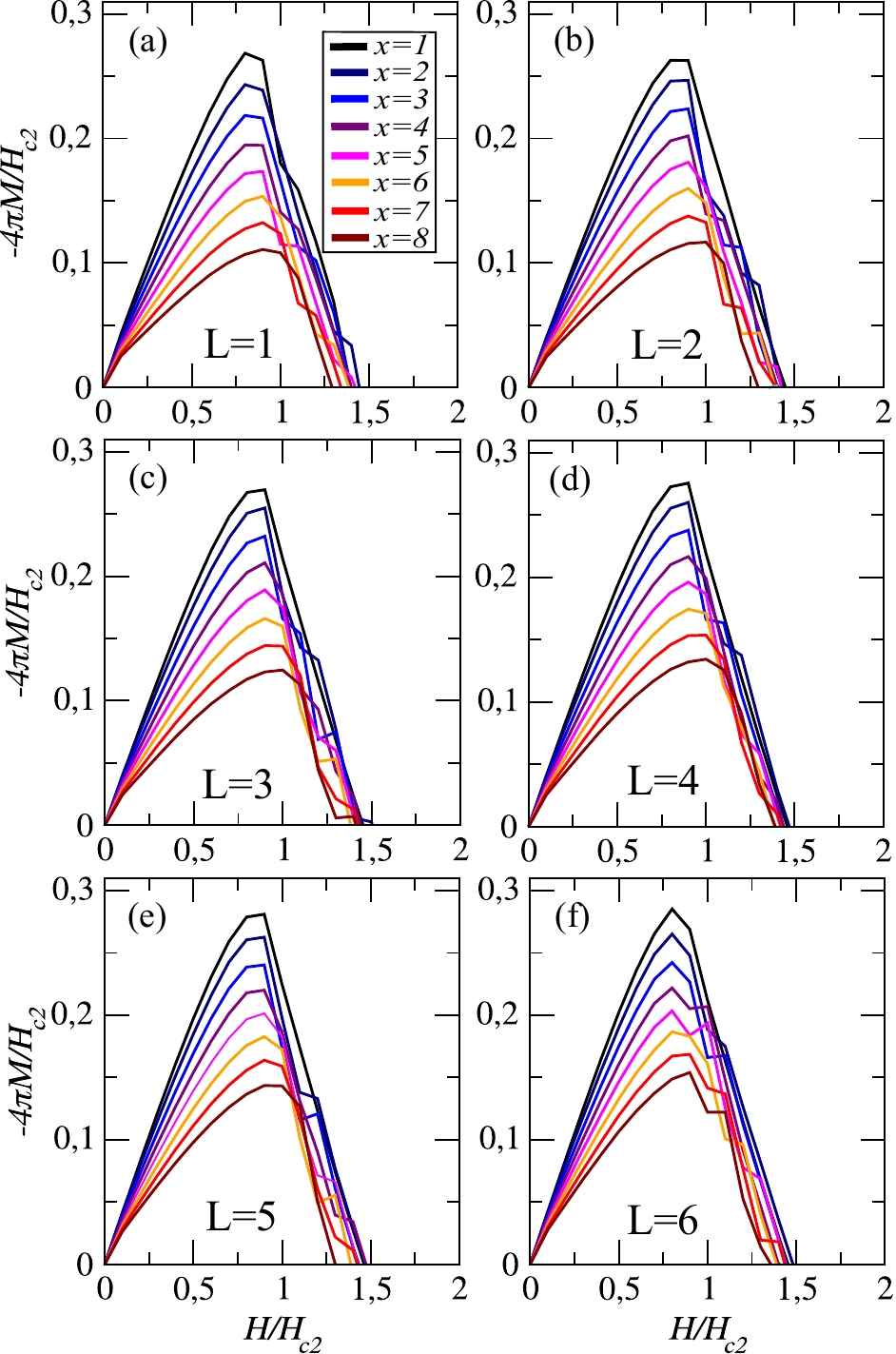}
    \caption{(a)-(f) Magnetization, $-4\pi\mathbf{M}/\mathbf{H}_{c2}$, as a function of the external magnetic field, $\mathbf{H}/\mathbf{H}_{c2}$, for  $1\leq \mathbf{x} \leq 8$ ($\Delta \mathbf{x}=1$), and $1 \leq \mathbf{L} \leq 6$ ($\Delta \mathbf{L}=1$).}
    \label{Magnetizacion2}
\end{figure}

\begin{figure}
    \centering
    \includegraphics[scale=0.677]{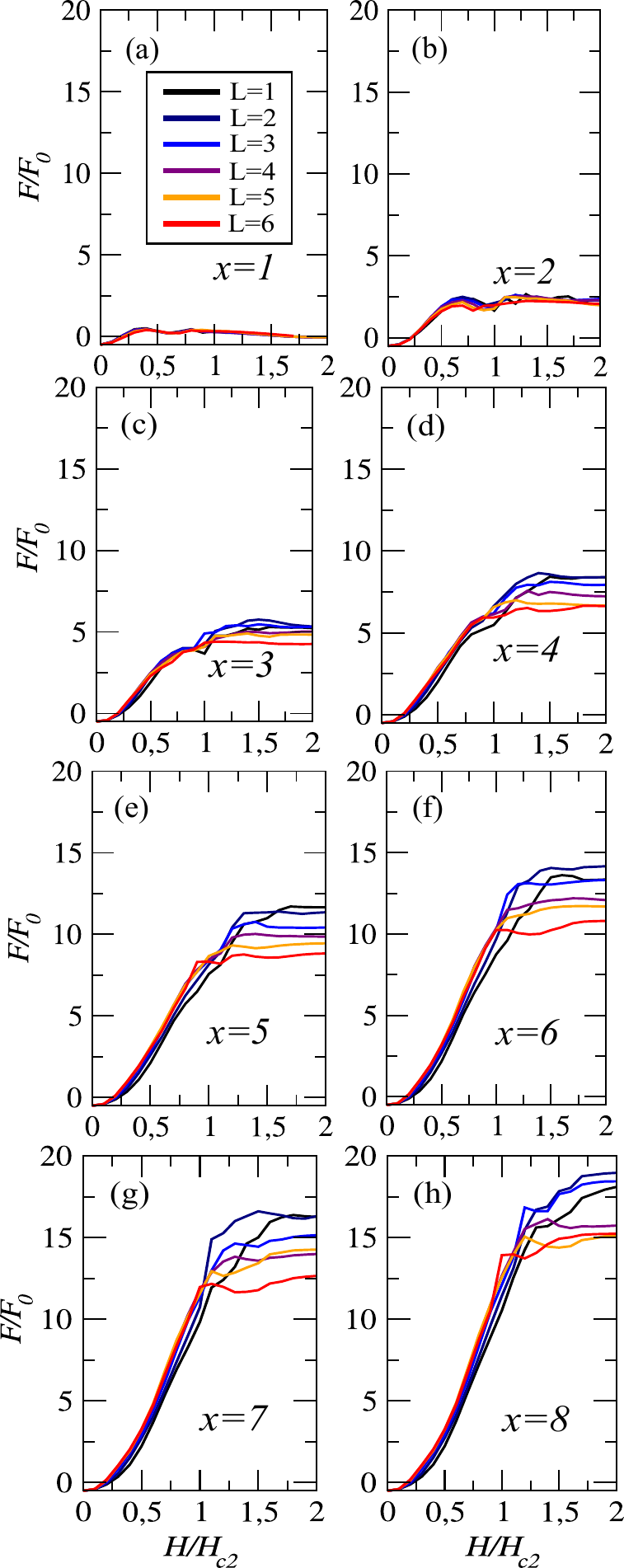}
    \caption{(a)-(h) Gibbs free energy density, $\mathbf{F}/\mathbf{F}_{0}$, as a function of  magnetic field, $\mathbf{H}/\mathbf{H}_{c2}$, for  $1\leq \mathbf{L} \leq 6$ ($\Delta \mathbf{L}=1$), and $1 \leq \mathbf{x} \leq 8$ ($\Delta \mathbf{x}=1$).}
    \label{Energy1}
\end{figure}

\begin{figure}
    \centering
    \includegraphics[scale=0.56]{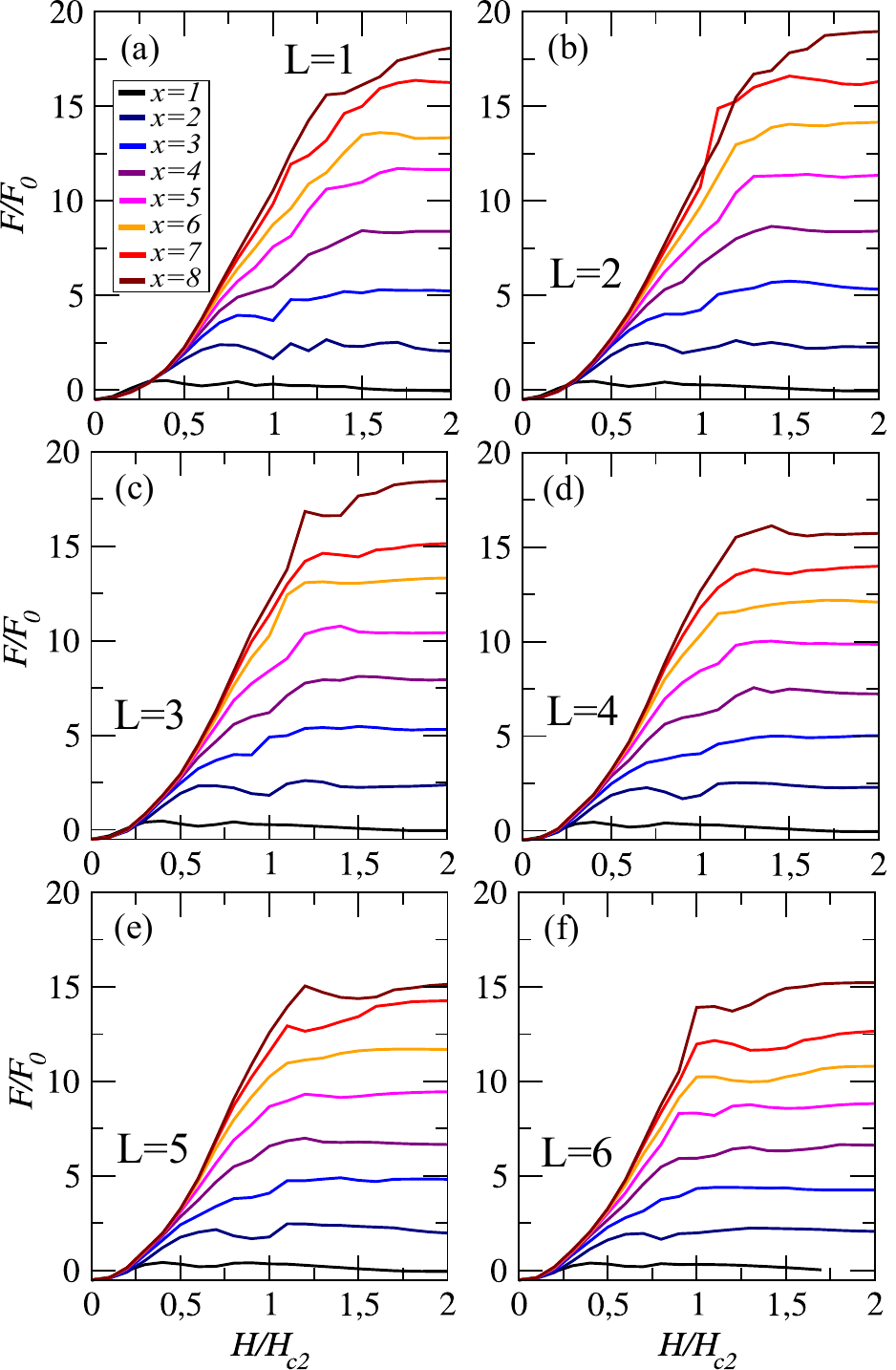}
    \caption{(a)-(f) Gibbs free energy density, $\mathbf{F}/\mathbf{F}_{0}$, as a function of magnetic field, $\mathbf{H}/\mathbf{H}_{c2}$, for  $1\leq \mathbf{x} \leq 8$ ($\Delta \mathbf{x}=1$), and $1 \leq \mathbf{L} \leq 6$ ($\Delta \mathbf{L}=1$).}
    \label{Energy2}
\end{figure}

\begin{figure*}
    \centering
  \includegraphics[scale=0.4]{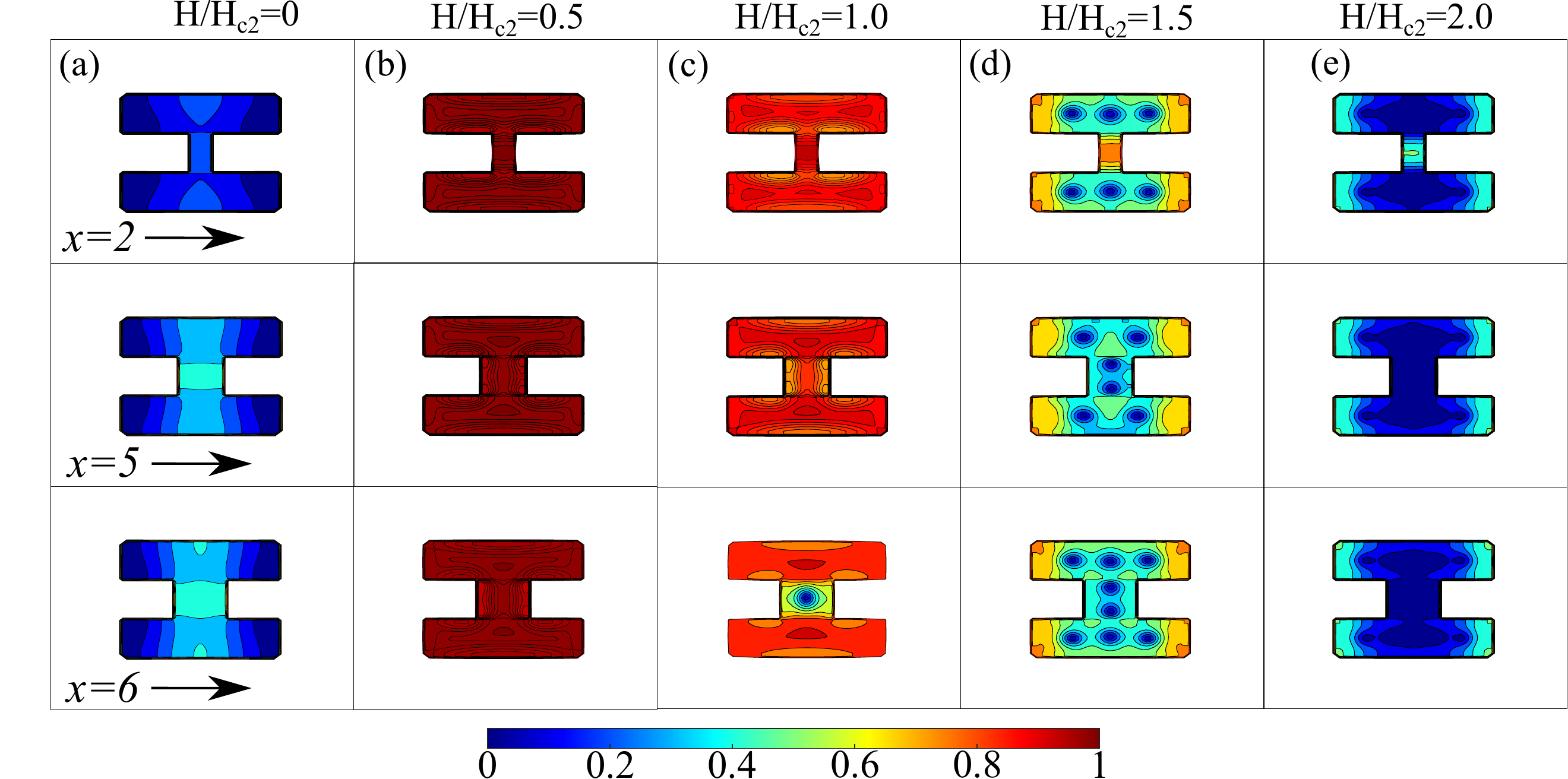}
    \caption{contour maps of Cooper pair density, $|\psi|^{2}$ for the superconducting nanobridge system, under  magnetic field $\mathbf{H}/\mathbf{H}_{c2}$, and for different values of  (a) $\mathbf{H}/\mathbf{H}_{c2}=0.0$, (b) $\mathbf{H}/\mathbf{H}_{c2}=0.5$, (c) $\mathbf{H}/\mathbf{H}_{c2}=1.0$, (d) $\mathbf{H}/\mathbf{H}_{c2}=1.5$ and (e) $\mathbf{H}/\mathbf{H}_{c2}=2.0$,  for a fixed $\mathbf{L}=5$ and $\mathbf{x}$.}
    \label{Psi1}
\end{figure*}

In Figs. \ref{Magnetizacion2}(a)-(f), we display $-4\pi \mathbf{M}/\mathbf{H}_{c2}$ as a function of $\mathbf{H}/\mathbf{H}_{c2}$ for fixed values of $1 \leq \mathbf{L} \leq 6$ with $\Delta\mathbf{L}=1$, and varying $1 \leq \mathbf{x} \leq 8$. We observe that for field values $(\mathbf{H}/\mathbf{H}_{c2} < 0.9)$, the sample is in the Meissner-Ochsenfeld state, as previously described, and $\mathbf{H}/\mathbf{H}_{c2}$ is shielded. However, the maximum value of $-4\pi \mathbf{M}/\mathbf{H}_{c2}$ decreases as $(\mathbf{x})$ increases, indicating the entry of vortex states into the superconducting system. This suggests that as the length of the nanobridge increases, more vortex states enter the system due to the greater number of superconducting states. For $\mathbf{H}/\mathbf{H}_{c2} > 1.5$, the system is in a normal state, i.e., $\mathbf{H}/\mathbf{H}_{c2} = 1.4$, which corresponds to the value of the second critical magnetic field, marking the limit for vortex state formation in the superconducting nanobridge system.\\\\\\
In Fig. \ref{Energy1}(a)-(h), we illustrate the Gibbs energy density $\mathbf{F}/\mathbf{F}_{0}$ as a function of $\mathbf{H}/\mathbf{H}_{c2}$ for a fixed $\mathbf{x}$ and varying $\mathbf{L}$. As $\mathbf{x}$ increases, $\mathbf{F}/\mathbf{F}_{0}$ generally increases, showing a maximum around $\mathbf{H}/\mathbf{H}_{c2} \sim 1 $ and remaining nearly constant for $\mathbf{H}/\mathbf{H}_{c2} > 1$. As expected, the minimum energy values occur for small $\mathbf{x}$ of the nanobridge and increase non-monotonically with the size of the nanobridge.\\\\
With this in mind, in Figs. \ref{Energy2}(a)-(f), we show $\mathbf{F}/\mathbf{F}_{0}$ as a function of $\mathbf{H}/\mathbf{H}_{c2}$ for fixed values of $\mathbf{L}$ and varying $\mathbf{x}$. Again, we observe that $\mathbf{F}/\mathbf{F}_{0}$ increases almost linearly with $\mathbf{H}/\mathbf{H}_{c2}$ and remains constant for $\mathbf{H}/\mathbf{H}_{c2} > 1$
This behavior of $\mathbf{F}/\mathbf{F}_{0}$ indicates that the fluxoids ($\mathbf{\Phi_{0}}$) in the superconducting system shift from their stable positions in the nanobridge and overlap due to the energetic interaction between the vortex states and the energy barrier at the boundaries. This effect is caused by the variation in $\mathbf{H}/\mathbf{H}_{c2}$. Consequently, we observe the limit for vortex state formation in the nanobridge.\\\\
In Fig. \ref{Psi1}, we display the Contour maps Cooper pair density $(\mathbf{|\psi|^{2}})$ for the superconducting nanobridge system at several values of $\mathbf{x}$ with a fixed $\mathbf{L}=5$, with the absence of $\mathbf{H}/\mathbf{H}_{c2}=0$ in (a), and under five values of $\mathbf{H}/\mathbf{H}_{c2}$ in (b) $\mathbf{H}/\mathbf{H}_{c2}=0.5$, (c) $\mathbf{H}/\mathbf{H}_{c2}=1.0$, (d) $\mathbf{H}/\mathbf{H}_{c2}=01.5$, and (e) $\mathbf{H}/\mathbf{H}_{c2}=2$. 
For $\mathbf{x}=2$, we observe that as $\mathbf{H}/\mathbf{H}_{c2}>0$ ($\mathbf{H}/\mathbf{H}_{c2}<2$) increases, vortex states form in both samples, with up to three vortex states in each sample at $\mathbf{H}/\mathbf{H}_{c2}=1.5$. However, the movement of vortex states in the nanobridge is not strongly evident. For larger values of $\mathbf{x}$, as $\mathbf{H}/\mathbf{H}_{c2}$ continues to increase, we observe the formation of two vortex states in each sample at $\mathbf{H}/\mathbf{H}_{c2}=1.5$, along with the movement of vortex states in the nanobridge, where two well-defined vortex states are shown. When $\mathbf{x}=6$, a single vortex state is observed in the center of the nanobridge at $\mathbf{H}/\mathbf{H}_{c2}=1$. As $\mathbf{H}/\mathbf{H}_{c2}$ continues to increase, three vortex states form in each sample, with the stabilization of two vortex states in the center of the nanobridge. Thus, we can affirm that the spatial dimensions of the nanobridge play a crucial role in controlling the vortex states within it.


\newpage
\subsection{Superconducting nanobridge system with induced currents}\label{sec3b}
In this section, we explore a second case involving a superconducting nanobridge system subjected to an induced $\mathbf{J}$, both positive and negative. $\mathbf{J}$ is introduced by varying the external magnetic field, $\mathbf{H}_{ext}/\mathbf{H}_{c2}$, at the systems boundaries, in accordance with the Ampere-Maxwell law. We examine the impact of this induced $\mathbf{J}$ under two specific conditions. First, we consider a condition where the $\mathbf{J}$ is applied only to one face of the single sample, as depicted in Fig. \ref{Layout}(b), and analyze how the increase in both internal and external $\mathbf{J}$ alters the vortex state within the superconducting nanobridge system. Subsequently, we investigate the behavior of vortex matter when $\mathbf{J}$ is introduced on both lateral faces of the superconducting nanobridge system.
\subsubsection{The current is induced on lateral face of single  sample} 
\begin{figure}
    \centering  \includegraphics[scale=0.52]{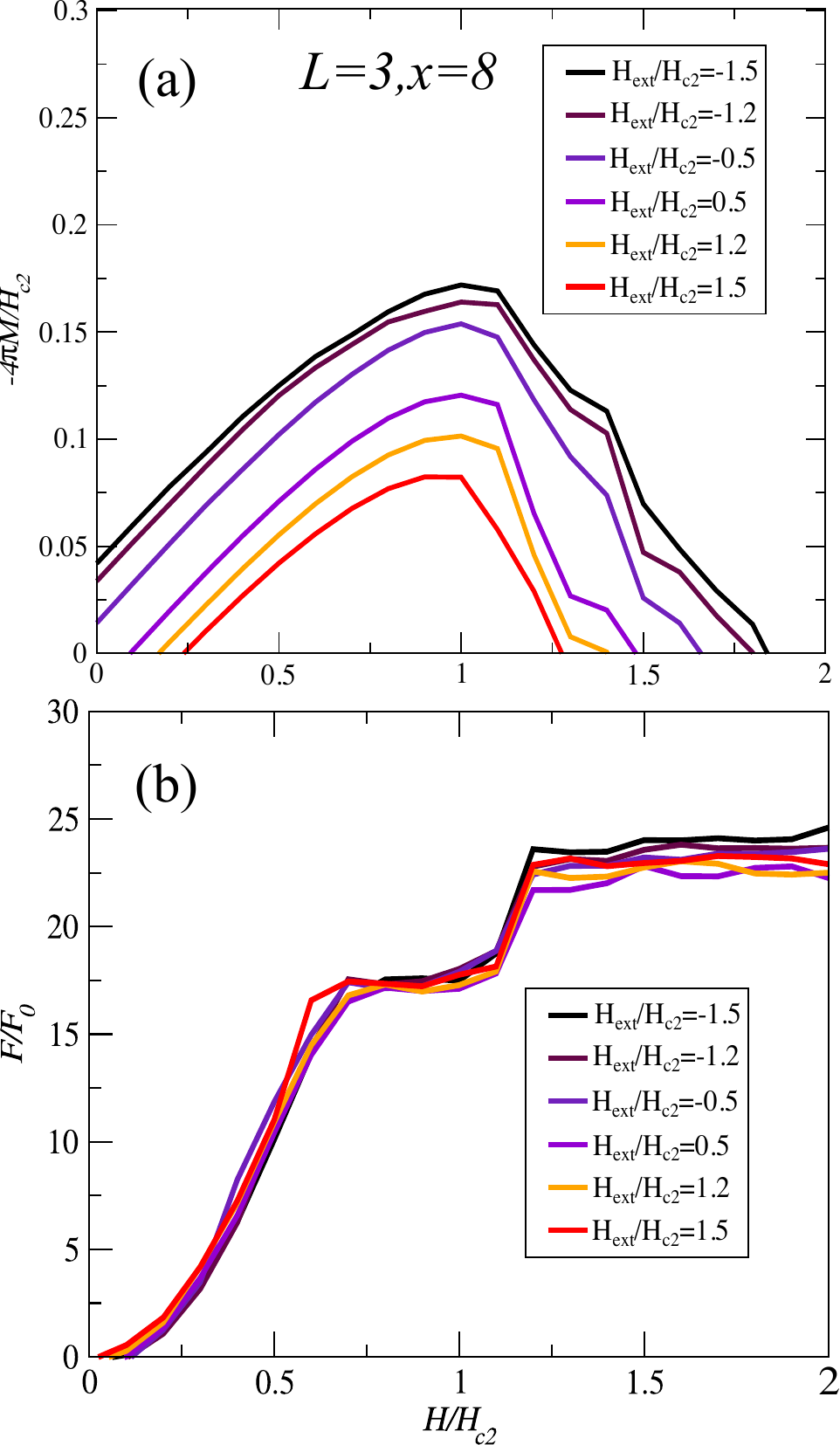}
    \caption{(a) Magnetization, ($-4\pi\mathbf{M}/\mathbf{H}_{c2}$) as a function of magnetic field $\mathbf{H}/\mathbf{H}_{c2}$ under different values of $\mathbf{H}_{ext}$ for fixed $\mathbf{L}$ and $\mathbf{x}$ values, and (b) Free Gibbs energy density,  $\mathbf{F}/\mathbf{F}_{0}$, as a function of magnetic field $\mathbf{H}/\mathbf{H}_{c2}$ under different values of $\mathbf{H}_{ext}$ for fixed $\mathbf{L}$ and $\mathbf{x}$ values.}
  \label{Enercurr}
\end{figure}

In Fig. \ref{Enercurr}(a), we show $-4\pi\mathbf{M}/\mathbf{H}_{c2}$ as a function of $\mathbf{H}/\mathbf{H}_{c2}$  under different values of $\mathbf{H}_{ext}/\mathbf{H}_{c2}$ (positive and negative) applying in the boundary of the superconducting nanobridge system, for fixed $\mathbf{L}$ and $\mathbf{x}$ values. We observe that for all values of $\mathbf{H}/\mathbf{H}_{c2} < 1.0$, the superconducting system remains in the Meissner-Oschenfeld state. For higher values of $\mathbf{H}_{ext}/\mathbf{H}_{c2}$, vortex states begin to penetrate the sample, leading to the initiation of the mixed state. Furthermore, as the induced $\mathbf{H}_{ext}/\mathbf{H}_{c2}$ at the boundaries of the superconducting sample increases, the mixed state is rapidly suppressed, accompanied by lower values of $-4\pi \mathbf{M}/\mathbf{H}_{c2}$. This behavior arises because the self-field generated in the superconducting nanobridge system, due to the boundary current, screens $\mathbf{H}_{ext}/\mathbf{H}_{c2}$ more efficiently, thereby minimizing the energy in the superconducting nanobridge system.\\\\

In Fig. \ref{Enercurr}(b), we illustrate $\mathbf{F}/\mathbf{F}_{0}$ as a function of $\mathbf{H}/\mathbf{H}_{c2}$ with a fixed induced 
$\mathbf{J}$ in the superconducting nanobridge system. We observe two significant jumps in 
$\mathbf{F}/\mathbf{F}_{0}$, where the energy barrier decreases, allowing vortex states to enter the lateral sections of the superconducting nanobridge system. Subsequently, an interaction is generated between the vortex states and the energy barrier at the boundaries, resulting in the non-monotonic behavior in 
$\mathbf{F}/\mathbf{F}_{0}$.\\\\
\begin{figure*}
    \centering
    \includegraphics[width=1\linewidth]{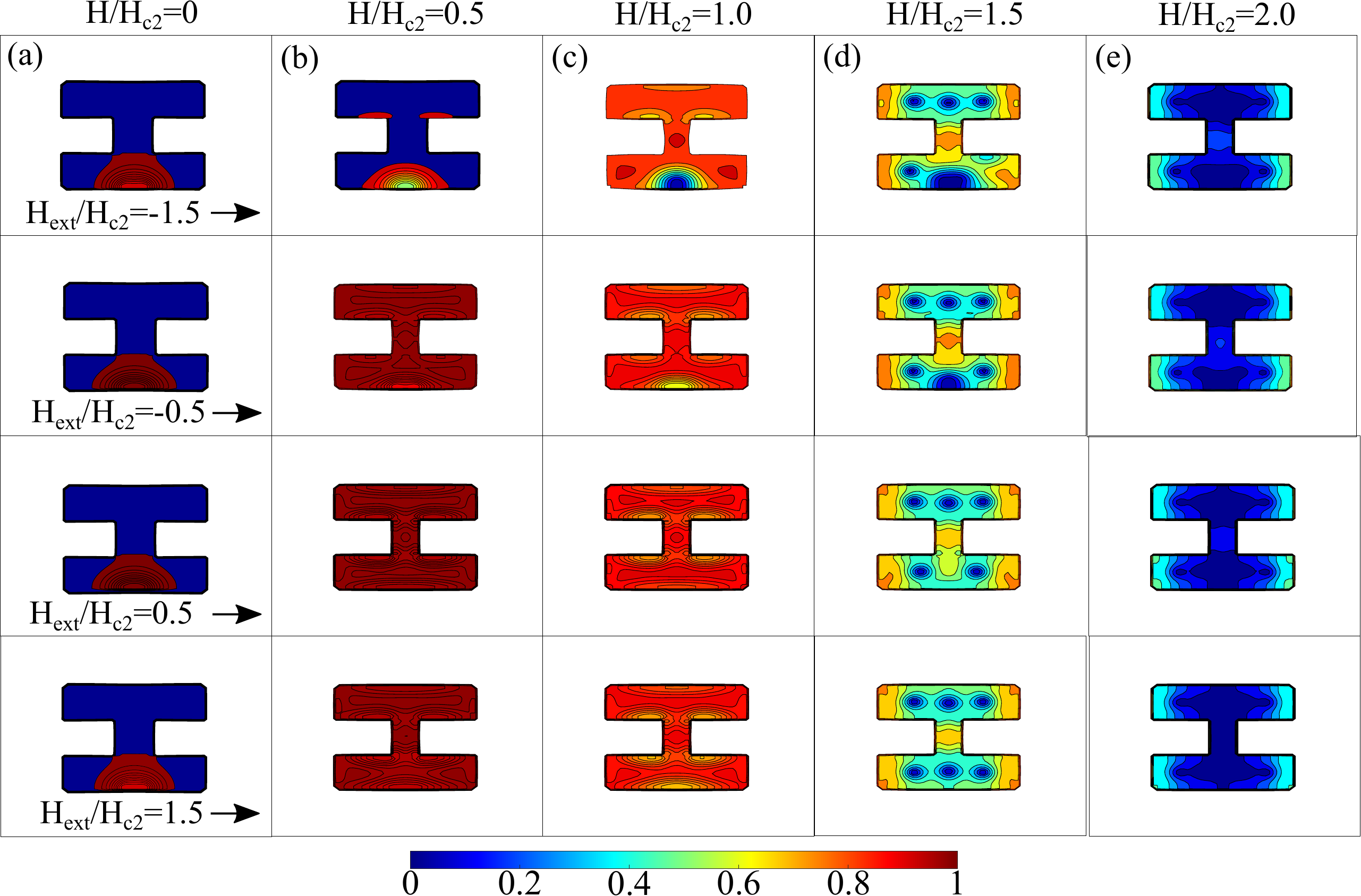}
    \caption{Contour maps of Cooper pair density, $|\psi|^{2}$ for the superconducting nanobridge system, under the induced magnetic field $\mathbf{H}_{ext}/\mathbf{H}_{c2}$, for different values of  (a) $\mathbf{H}/\mathbf{H}_{c2}=0.0$, (b)$\mathbf{H}/\mathbf{H}_{c2}=0.5$, (c) $\mathbf{H/\mathbf{H}_{c2}}=1$, (d) $\mathbf{H}/\mathbf{H}_{c2}=1.5$, and (e) $\mathbf{H}/\mathbf{H}_{c2}=2$  for $\mathbf{L}=5$ and $\mathbf{x}=$ 2, 5 and 6.}
    \label{Psi2}
\end{figure*}
To analyse the vortex states configuration, when just one of face of the superconducting nanobridge system has a induced external $\mathbf{J}$ in Fig. \ref{Psi2}(a)-(e), we display the contour maps of density of Cooper pairs, for different values of $\mathbf{H}/\mathbf{H}_{c2}$ and induced $\mathbf{H}_{ext}/\mathbf{H}_{c2}$ (positive and negative). 
We observe that for $\mathbf{H}/\mathbf{H}_{c2}=0$, regardless of the intensity of $\mathbf{H}_{ext}/\mathbf{H}_{c2}$, the mere presence of one of the faces of the superconducting nanobridge system with an induced $\mathbf{J}$ already generates vortex states at the contact. As $\mathbf{H}/\mathbf{H}_{c2}$ increases, it becomes evident that for all values of $\mathbf{H}_{ext}/\mathbf{H}_{c2}$, the stabilization of vortex states becomes more pronounced, as seen at $\mathbf{H}/\mathbf{H}_{c2}=1.5$, where vortex states appear asymmetrically on the two sides of the superconducting nanobridge system. Then, we can deduce that the induction of $\mathbf{J}$ on one face of the superconducting nanobridge system restricts the passage of vortex states through the nanobridge.\\\\
\subsubsection{The induced current is on both lateral faces of superconducting samples} 
\begin{figure}
    \centering
    \includegraphics[width=0.96\linewidth]{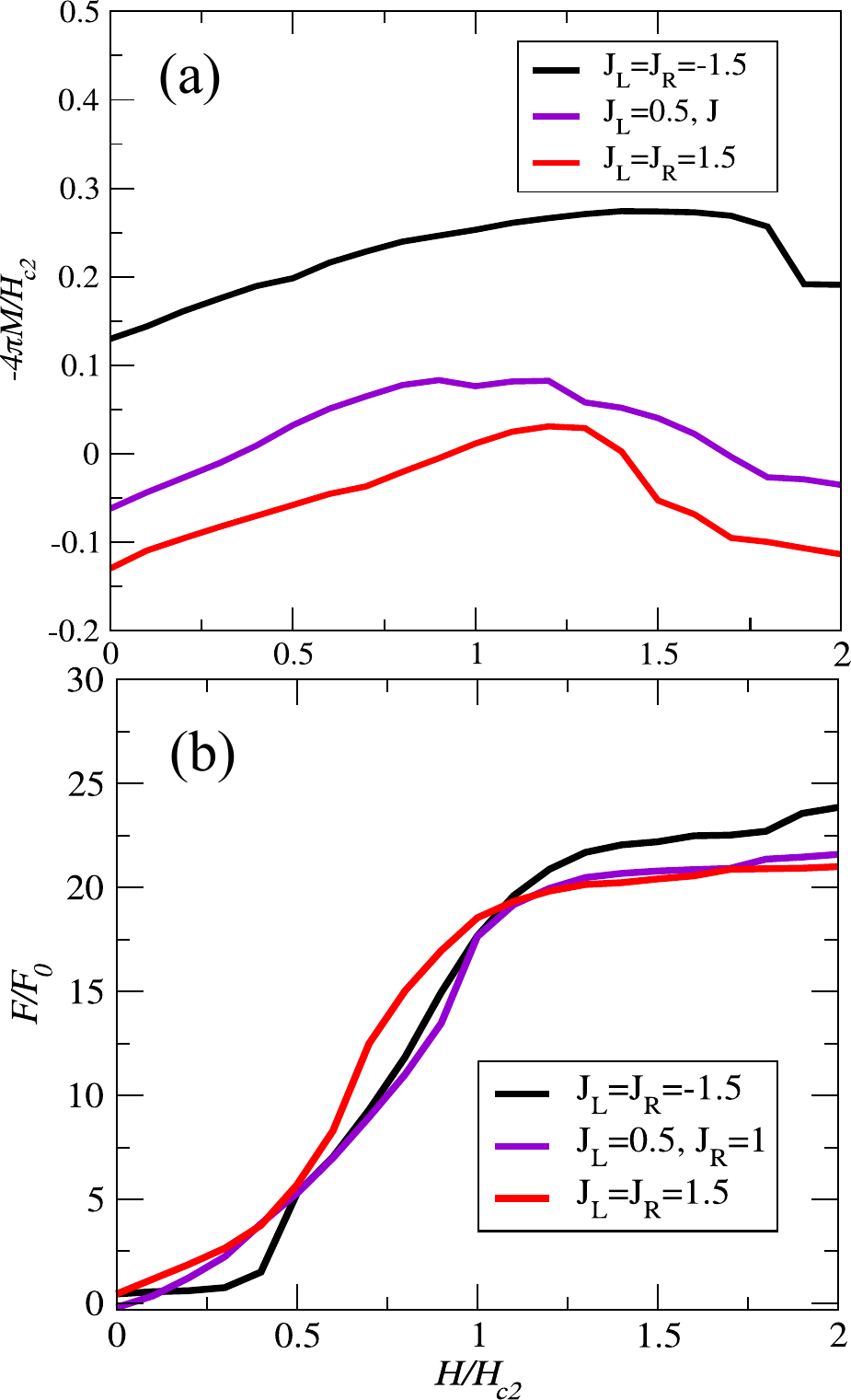}
    \caption{(a) Magnetization $-4\pi\mathbf{M}/\mathbf{H}_{c2}$ and (b) Gibbs free energy density as a functions of the magnetic field, $\mathbf{H}/\mathbf{H}_{c2}$, for three different values of external current, $\mathbf{J}
    $.}
    \label{figM2C2}
\end{figure}

\begin{figure*}
    \centering
    \includegraphics[width=1\linewidth]{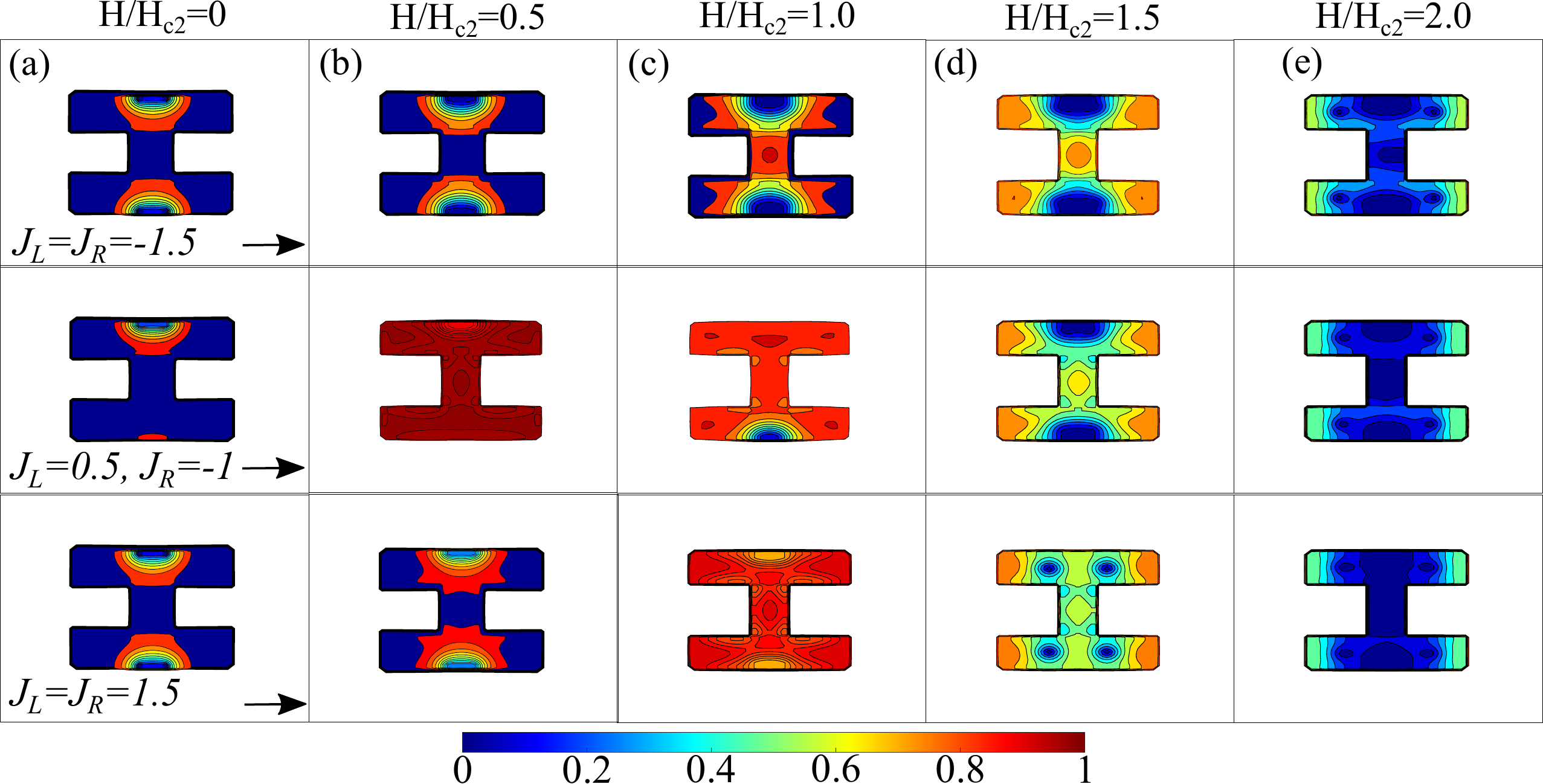}
    \caption{Contour maps of Cooper pair density, $|\psi|^{2}$ of the superconducting nanobridge system, under magnetic field $\mathbf{H}/\mathbf{H}_{c2}$, when  (a) $\mathbf{H}/\mathbf{H}_{c2}=0.0$, (b) $\mathbf{H}/\mathbf{H}_{c2}=0.5$, (c) $\mathbf{H}/\mathbf{H}_{c2}=1.0$, (d) $\mathbf{H}/\mathbf{H}_{c2}=1.5$ and (e) $\mathbf{H}/\mathbf{H}_{c2}=2.0$ for $\mathbf{L}=5$ and different values of $\mathbf{x}$.}
    \label{Psi2c}
\end{figure*}

Based on the previous results, we investigate the effects in the nanobridge under the influence of a $\mathbf{J}$ applied to both faces of the superconducting nanobridge system, as illustrated in Fig. \ref{Layout}(b). For this analysis, we consider fixed values of $\mathbf{L}=3$ and $\mathbf{x}=8$ while varying the intensity of the induced external $\mathbf{J}$. Fig. \ref{figM2C2} displays the behavior of $-4\pi \mathbf{M}/\mathbf{H}_{c2}$ as a function of $\mathbf{H}/\mathbf{H}_{c2}$ for three different values of the induced $\mathbf{J}$, denoted as $\mathbf{J}_{L}$ and $\mathbf{J}_{R}$, respectively. These currents, $\mathbf{J}_{L}$ and $\mathbf{J}_{R}$, are labeled R and L to signify their application on the left and right faces of the superconducting nanobridge system. It is observed that $-4\pi \mathbf{M}/\mathbf{H}_{c2}$ increases as the intensity of the induced $-\mathbf{H}/\mathbf{H}_{c2}$ increases, reaching a maximum before subsequently decaying. This indicates that the system achieves a state where $-4\pi \mathbf{M}/\mathbf{H}{c2}$ is maximized. In addition, it is evident that the behavior of $-4\pi \mathbf{M}/\mathbf{H}{c2}$ varies with different intensities of the applied $\mathbf{J}$, leading to distinct vortex state configurations in the superconducting nanobridge system.\\\\\\
In Fig. \ref{figM2C2}(b), we present $\mathbf{F}/\mathbf{F}_{0}$ as a function of $\mathbf{H}/\mathbf{H}_{c2}$ for the induced $\mathbf{J}$ in the superconducting nanobridge system. Two notable jumps in $\mathbf{F}/\mathbf{F}_{0}$ are observed, corresponding to decreases in the energy barrier that facilitate the entry of vortex states into the lateral sections of the sample. Subsequently, an interaction arises between the vortex states and the energy barrier at the boundaries, resulting in the non-monotonic behavior of $\mathbf{F}/\mathbf{F}_{0}$.
Furthermore, to elucidate the configuration of vortex states when a current $\mathbf{J}$ is induced on both faces of the superconducting nanobridge system, we illustrate the contour maps of the Cooper pair densities in Fig. \ref{Psi2c}(a)-(e) for various values of $\mathbf{H}/\mathbf{H}_{c2}$, where $0 \leq \mathbf{H}/\mathbf{H}_{c2} \leq 2$ with increments of $\Delta \mathbf{H}/\mathbf{H}_{c2} = 0.5$. For $\mathbf{J}_{L} = \mathbf{J}_{R} = -1.5$, we observe that at both $\mathbf{H}/\mathbf{H}_{c2} = 0$ and $\mathbf{H}/\mathbf{H}_{c2} = 0.5$, vortex states appear on both faces of the superconducting nanobridge system, indicating the presence of vortex states even with the absence of $\mathbf{J}$. In Fig. \ref{figM2C2}(a)  $-4\pi \mathbf{M}/\mathbf{H}_{c2}$ is non-zero at $\mathbf{H}/\mathbf{H}_{c2}=0$. For $\mathbf{H}/\mathbf{H}_{c2} > 1.0$, we observe the formation of vortex states in the center of the superconducting system, corresponding to the point of greatest symmetry in the nanobridge, specifically characterized by the formation of a single fluxoid.\\\\\\
In the superconducting nanobridge system, when $\mathbf{H}/\mathbf{H}_{c2} > 1$, five vortex states can be observed: two in each lateral section of the sample and a single vortex state at the center of the nanobridge. In the same context, when $\mathbf{J}_{L} = 0.5$ and $\mathbf{J}_{R} = -1$ at $\mathbf{H}/\mathbf{H}_{c2} = 0$, it is evident that the induced $\mathbf{J}$ at the contacts naturally produce nucleation of vortex states. Due to the difference in $\mathbf{J}$ intensities, the vortex states exhibit varying intensities. As $\mathbf{H}/\mathbf{H}_{c2}$ increases, vortex states begin to nucleate at the center of the nanobridge, where the cross section is located, leading to the formation of a single fluxoid. Additionally, when $\mathbf{J}$ is induced in different directions, such as with $\mathbf{J}_{L} = \mathbf{J}_{R} = 1.5$, we observe that even in the absence of an external $\mathbf{H}/\mathbf{H}_{c2}$ fi, the mere presence of $\mathbf{J}$ applied to both faces of the superconducting nanobridge system is sufficient to generate the entry of vortex states. As $\mathbf{H}/\mathbf{H}_{c2}$ further increases, the nucleation of vortex states becomes more pronounced at the center of the nanobridge, driven by the current $\mathbf{J}$ connected to both the left and right faces of the superconducting nanobridge system.

\section{Conclusions}\label{sec4}
In this work, we have investigated various electronic variables such as magnetization, Gibbs free energy, and Cooper pair density in a superconducting  system composed of two samples connected by a nanobridge. We study two specific cases of interest. The first case examines the limit where the length and/or thickness of the nanobridge is sufficient to generate vortex states. In this case, we have observed how spatial dimensions influence the density of Cooper pairs, particularly at low magnetic field intensities. The second case explores the effects produced in the superconducting nanobridge system due to the application of an external current. We discuss two conditions: the first where current is induced on a single face of the superconducting nanobridge system, and the second where current is applied to both faces of the system. With this in mind, we have derived several important insights.\\\\
\begin{itemize}
\item (i) in the superconducting nanobridge system, in the absence of induced external currents, vortex nucleation occurs, and these vortex states move through the nanobridge connecting the two lateral samples. Consequently, by modifying the spatial dimensions of the nanobridge, defined by $\mathbf{L}$ and $\mathbf{x}$, it becomes evident that these dimensions play a crucial role in the configuration of vortex states within the superconducting nanobridge system.\\
\item (ii) With the application of external currents on only one lateral face of the superconducting nanobridge system, vortex nucleation occurs as the magnetic field increases. However, this phenomenon is observed only within the samples, and no vortex state movement occurs across the nanobridge.\\\\ 
\item (iii) by inducing external currents on both faces of the superconducting nanobridge system, vortex nucleation can occur, with vortices moving through the nanobridge, regardless of the bridge's spatial dimensions. This holds true even if the currents on the two faces are of different intensities. This suggests that vortex nucleation or transport through the nanobridge will occur whenever currents are induced on both faces of the superconducting nanobridge system.\\
\end{itemize}
Our theoretical results can open the doors for technological studies, such as the development of efficient memories \cite{Kalashnikov2023}, where the presence of vortex states in the nano/bridge can represent bits [0,1] or development of quantum logic operations through topological \textit{qubits} \cite{Demler2018,Metlitski2017,Singh2021}. Finally, the vortex states in a nanobridge can be manipulated to detect particles or radiation (such as single photons), where the interaction with the radiation can interact with the vortex states in a detectable way, among other possible uses and developments of the new technological era \cite{Zolotov2014,Levchenko2020}.  
\section{ACKNOWLEDGMENTS}
 \textbf{C. Aguirre} thanks \textbf{S. Aguirre} and \textbf{M. Aguirre} for useful discussions.  \textbf{C. Aguirre}  also thanks to \textbf{Cnpq} grant number process: 174045/2023-9 for financial support.
\textbf{J. Faúndez} thanks the \textbf{FAPERJ} (Fundação de Amparo à Pesquisa do Estado do Rio de Janeiro), process n° SEI-260003/019642/2022, and the partial support from \textbf{ANID} Fondecyt grant number 3240320. \textbf{P.D.} and \textbf{D.L.} acknowledge partial financial support from \textbf{FONDECYT} 1231020. \textbf{J. Barba-Ortega} thanks Alejandro and Marcos for useful discussions. This research was partially supported by the supercomputing infrastructure of the NLHPC (CCSS210001).


\end{document}